\documentclass[3p,times,twocolumn]{elsarticle}
\usepackage{amssymb}
\usepackage{color}

\biboptions{numbers,sort&compress}

\makeatletter
\def\ps@pprintTitle{%
     \let\@oddhead\@empty
     \let\@evenhead\@empty
     \let\@oddfoot\@empty
     \let\@evenfoot\@empty}
\makeatother

\newcommand{\DP}{\Delta\Pi}
\newcommand{\ind}[2]{^{\mbox{\scriptsize $#1$}}_{\mbox{\scriptsize #2}}}
\newcommand{\inds}[2]{^{\mbox{\scriptsize $#1$}}_{\mbox{\tiny #2}}}

\newcommand{\nf}{n_{\mbox{\scriptsize f}}}

\newcommand\AL[1]{\Delta\alpha\ind{#1}{lep}}
\newcommand\AH[1]{\Delta\alpha\ind{#1}{had}}
\newcommand\AMH[1]{\Delta\alpha\ind{#1}{had}}
\newcommand{\txt}[2]{\color{#2}\scriptsize\textsf{#1}}
\definecolor{G4}{rgb}{0.00,0.55,0.00}
\definecolor{RB4}{rgb}{0.15,0.25,0.55}

\begin{document}

\begin{frontmatter}
\title{Hadronic contributions to electroweak observables within~DPT}
\author{A.V.~Nesterenko}
\ead{nesterav@theor.jinr.ru}
\address{Bogoliubov Laboratory of Theoretical Physics,
Joint Institute for Nuclear Research,
Dubna, 141980, Russian Federation}
\begin{abstract}
Dispersive approach to quantum chromodynamics is applied to the assessment
of hadronic contributions to electroweak observables. The~employed
approach merges the corresponding perturbative input with intrinsically
nonperturbative constraints, which originate in the respective kinematic
restrictions. The evaluated hadronic contributions to the muon anomalous
magnetic moment and to the shift of the electromagnetic fine structure
constant at the scale of $Z$~boson mass conform with recent assessments of
these quantities.
\end{abstract}
\end{frontmatter}

Numerous strong interaction processes are governed by the hadronic vacuum
polarization function~$\Pi(q^2)$. Its~ultraviolet behavior can be studied
within perturbation theory, whereas its infrared behavior is only
accessible within various nonperturbative approaches, e.g., lattice
simulations~\cite{LattRev, Lat1, Lat2, Lat3, Lat4a, Lat4b}, operator
product expansion~\cite{OPE1, OPE23, OPE4, OPE5, OPE6}, instanton liquid
model~\cite{NLCQM, ILM}, and others.

Certain nonperturbative information about the low--energy hadron dynamics
is contained within dispersion relations. The latter are widely employed
in various issues of theoretical particle physics, for example, the
precise determination of parameters of resonances~\cite{Kaminski}, the
extension of applicability range of chiral perturbation
theory~\cite{Portoles, Passemar}, the assessment of the hadronic
light--by--light scattering~\cite{DispHlbl}, and many others (see, e.g.,
Refs.~\cite{APT, APT1, APT2, APT3, APT4, APT5a, APT5b, APT6, APT7a, APT7b,
APT8, APT9, APT10, APT11}).

The dispersion relations render the kinematic restrictions on the relevant
physical processes into the mathematical form and impose stringent
intrinsically nonperturbative constraints on the pertinent quantities.
Among the latter are the function~$\Pi(q^2)$, which is defined as the
scalar part of the hadronic vacuum polarization tensor
\vspace*{-1.5mm}
\begin{eqnarray}
\label{P_Def}
\Pi_{\mu\nu}(q^2) \!\!\!\!&=&\!\!\!\! i\!\int\!d^4x\,e^{i q x} \bigl\langle 0 \bigl|
T\bigl\{J_{\mu}(x)\, J_{\nu}(0)\bigr\} \bigr| 0 \bigr\rangle = \qquad
\nonumber \\[-0.5mm]
&=&\!\!\!\! i\, (q_{\mu}q_{\nu} - g_{\mu\nu}q^2)\, \Pi(q^2)/(12\pi^2),
\end{eqnarray}
related $R(s)$~function
\begin{equation}
\label{R_Def}
R(s) = \rm{Im}\lim_{\varepsilon \to 0_{+}}\!\Pi(s+i\varepsilon)/\pi,
\end{equation}
which is identified with the $R$--ratio of electron--positron annihilation
into hadrons, and the Adler function~\cite{Adler}
\begin{equation}
\label{Adler_Def}
D(Q^2) = - \frac{d\, \Pi(-Q^2)}{d \ln Q^2},
\end{equation}
with $Q^2 = -q^2 > 0$ and $s = q^2 > 0$ being the spacelike and timelike
kinematic variables, respectively.

The dispersive approach to QCD~\cite{DPT1a, PRD88, JPG42} (its preliminary
formulation was discussed in Refs.~\cite{DPTPrelim1, DPTPrelim2}) merges
the aforementioned nonperturbative constraints with corresponding
perturbative input and provides the unified integral representations for
the functions on hand:
\begin{eqnarray}
\label{P_DPT}
&&\hspace{-12mm}
\DP(q^2,\, q_0^2) = \DP^{(0)}(q^2,\, q_0^2) +
\nonumber \\
&&+\int_{m^2}^{\infty} \rho(\sigma)
\ln\biggl(\frac{\sigma-q^2}{\sigma-q_0^2}
\frac{m^2-q_0^2}{m^2-q^2}\biggr)\frac{d\,\sigma}{\sigma}, \\
\label{R_DPT}
&&\hspace{-12mm}
R(s) = R^{(0)}(s) + \theta(s-m^2) \int_{s}^{\infty}\!
\rho(\sigma) \frac{d\,\sigma}{\sigma}, \\
\label{Adler_DPT}
&&\hspace{-12mm}
D(Q^2) = D^{(0)}(Q^2) +
\nonumber \\
&&+\frac{Q^2}{Q^2+m^2}
\int_{m^2}^{\infty} \rho(\sigma)
\frac{\sigma-m^2}{\sigma+Q^2} \frac{d\,\sigma}{\sigma}.
\end{eqnarray}
Here $\DP(q^2\!,\, q_0^2) = \Pi(q^2) - \Pi(q_0^2)$, $m^2 = 4m_{\pi}^2$,
$\theta(x)$~is the unit step--function [$\theta(x)=1$ if $x \ge 0$ and
$\theta(x)=0$ otherwise], the leading--order terms read~\cite{Feynman,
QEDAB}
\begin{eqnarray}
\label{P0L}
&&\hspace{-12mm}
\DP^{(0)}(q^2,\, q_0^2) = 2\,\frac{\varphi - \tan\varphi}{\tan^3\varphi}
- 2\,\frac{\varphi_{0} - \tan\varphi_{0}}{\tan^3\varphi_{0}}, \\
\label{R0L}
&&\hspace{-12mm}
R^{(0)}(s) = \theta(s - m^2)\bigl[1-(m^2/s)\bigr]^{3/2}, \\
\label{D0L}
&&\hspace{-12mm}
D^{(0)}(Q^2) = 1 + 3\Bigl[1 - \sqrt{1 + \xi^{-1}}\,
\sinh^{-1}\bigl(\xi^{1/2}\bigr)\Bigr]\xi^{-1},
\end{eqnarray}
and $\rho(\sigma)$ stands for the spectral density
\begin{eqnarray}
\label{RhoGen}
&&\hspace*{-12mm}
\rho(\sigma) = \frac{1}{\pi} \frac{d}{d\,\ln\sigma}\,
\mbox{Im}\lim_{\varepsilon \to 0_{+}} p(\sigma-i\varepsilon) =
- \frac{d\, r(\sigma)}{d\,\ln\sigma} =
\nonumber \\[2mm]
&&\hspace*{-4.5mm}
= \rm{Im}\lim_{\varepsilon \to 0_{+}} d(-\sigma-i\varepsilon)/\pi.
\end{eqnarray}
In these equations $\sin^2\!\varphi = q^2/m^2$, $\sin^2\!\varphi_{0} =
q^{2}_{0}/m^2$, $\xi=Q^2/m^2$, and $p(q^2)$, $r(s)$, $d(Q^2)$
denote the strong corrections to the respective functions, see
Refs.~\cite{DPT1a, PRD88, JPG42} for the details.
\looseness=-1

It~is worth noting that the
representations~(\ref{P_DPT})--(\ref{Adler_DPT}) conform with the results
of Bethe--Salpeter calculations~\cite{PRL99PRD77} as well as of lattice
simulations~\cite{RCTaylor}. The Adler function~(\ref{Adler_DPT}) agrees
with its experimental prediction in the entire energy range~\cite{DPT1a,
DPT1b, DPT2} (the studies of~$D(Q^2)$ within other approaches can be found
in, e.g., Refs.~\cite{MSS, Cvetic, Maxwell, Kataev, Fischer, PeRa, BJ}).
Additionally, the dispersive approach has proved to be capable of
describing OPAL (update~2012, Ref.~\cite{OPAL9912}) and ALEPH
(update~2014, Ref.~\cite{ALEPH0514}) experimental data on inclusive
$\tau$~lepton hadronic decay in vector and axial--vector channels in a
self--consistent way~\cite{PRD88, QCD14} (see also Refs.~\cite{DPT3,
C12}).

The perturbative part of the spectral density can be calculated as (see,
e.g., Refs.~\cite{CPC, BCmath})
\begin{eqnarray}
\label{RhoPert}
&&\hspace*{-12mm}
\rho\ind{}{pert}(\sigma) = \frac{1}{\pi} \frac{d}{d\,\ln\sigma}\,
\mbox{Im}\lim_{\varepsilon \to 0_{+}} p\ind{}{pert}(\sigma-i\varepsilon) =
\nonumber \\
&&\hspace*{-5mm}
= \! - \frac{d\, r\ind{}{pert}(\sigma)}{d\,\ln\sigma}
\!=\! \frac{1}{\pi}\, \mbox{Im}\lim_{\varepsilon \to 0_{+}}
d\ind{}{pert}(-\sigma-i\varepsilon),
\end{eqnarray}
that provides the respective perturbative input to the integral
representations (\ref{P_DPT})--(\ref{Adler_DPT}). The latter are by
construction consistent with aforementioned nonperturbative constraints
and corresponding perturbative results and constitute the ``dispersively
improved perturbation theory''~(DPT) expressions for the functions on
hand. At the one--loop level Eq.~(\ref{RhoPert}) assumes a quite simple
form, namely, $\rho\ind{(1)}{pert}(\sigma) = (4/\beta_{0})
[\ln^{2}(\sigma/\Lambda^2)+\pi^2]^{-1}$, where $\beta_{0} = 11 - 2\nf/3$,
$\nf$~denotes the number of active flavors, and~$\Lambda$ is the QCD scale
parameter. The explicit expressions for the spectral
function~(\ref{RhoPert}) up to the four--loop level are given in
Ref.~\cite{CPC} (recently calculated respective four--loop perturbative
coefficient can be found in Ref.~\cite{AdlerPert4L}). The perturbative
spectral function~(\ref{RhoPert}) will be employed hereinafter.
\looseness=-1

Note that in the massless limit $(m=0)$ for the case of perturbative
spectral function~(\ref{RhoPert}) Eqs.~(\ref{R_DPT}) and~(\ref{Adler_DPT})
become identical to those of the ``analytic perturbation
theory''~(APT)~\cite{APT} (see also Refs.~\cite{APT1, APT2, APT3, APT4,
APT5a, APT5b, APT6, APT7a, APT7b, APT8, APT9, APT10, APT11}). However, as
discussed in Refs.~\cite{PRD88, DPT1a, JPG42, C12, DPT2}, the massless
limit loses the substantial nonperturbative constraints, which relevant
dispersion relations impose on the functions on hand, that appears to be
essential for the studies of hadron dynamics at low energies.

\begin{figure}[t]
\centerline{\includegraphics[width=75mm]{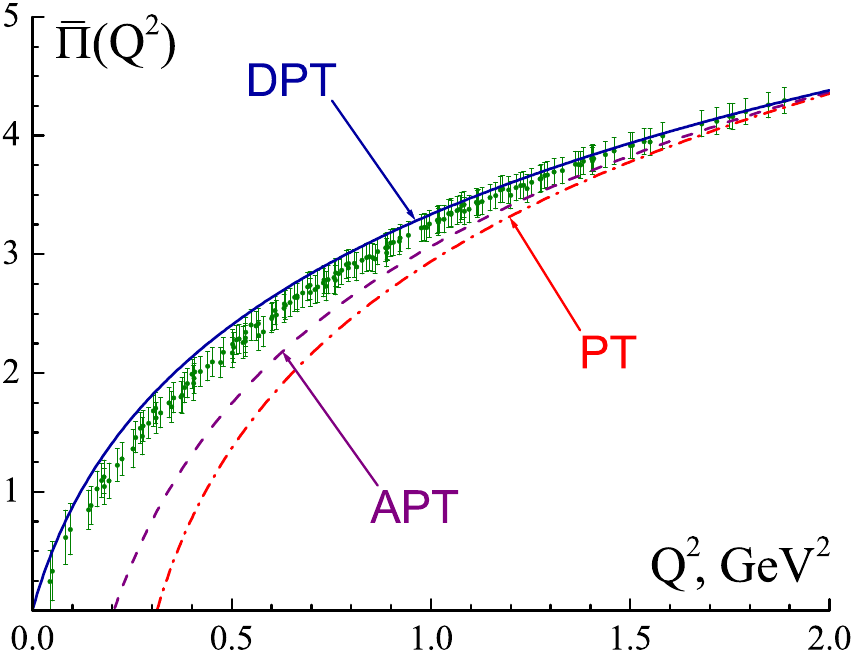}}
\caption{Hadronic vacuum polarization function within various approaches:
DPT~[Eq.~(\ref{P_DPT2}), solid curve], APT~[Eq.~(\ref{PAPT1L}), dashed
curve], PT~[Eq.~(\ref{PPert1L}), dot--dashed curve], and lattice
data~(Ref.~\cite{Lat5}, circles).}
\label{Plot:P_DPT}
\end{figure}

In what follows it is convenient to employ the subtracted at zero form of
Eq.~(\ref{P_DPT}), specifically
\begin{eqnarray}
\label{P_DPT2}
&&\hspace{-12mm}
\bar{\Pi}(Q^2) = \DP(0,-Q^2) = \DP^{(0)}(0, -Q^2) +
\nonumber \\
&& \hspace{0.9mm}
+ \int_{m^2}^{\infty} \rho(\sigma)
\ln\biggl(\frac{1+Q^2/m^2}{1+Q^2/\sigma}\biggr)
\frac{d\,\sigma}{\sigma}.
\end{eqnarray}
As one can infer from Fig.~\ref{Plot:P_DPT}, the obtained hadronic vacuum
polarization function (solid curve) is in a good agreement with lattice
data~\cite{Lat5}~(circles) (the rescaling procedure described in
Refs.~\cite{RhoRescale1, RhoRescale2} was applied). The presented result
corresponds to the four--loop level, $\Lambda=419\,$MeV, and~$\nf=2$.
Figure~\ref{Plot:P_DPT} also displays the one--loop Eq.~(\ref{P_DPT}) in
the massless limit, which corresponds to APT (dashed curve)
\begin{equation}
\label{PAPT1L}
\hspace*{-6.5mm}
\DP\inds{(1)}{APT}(-Q_{0}^{2},-Q^{2}) = \ln\biggl(\frac{Q^2}{Q_0^2}\biggr)
+ \frac{4}{\beta_{0}}
\ln\Biggl[\frac{a\ind{(1)}{an}(Q_{0}^{2})}{a\ind{(1)}{an}(Q^{2})}\Biggr],
\end{equation}
and the one--loop perturbative approximation of~$\Pi(q^2)$ (dot--dashed
curve)
\begin{equation}
\label{PPert1L}
\hspace*{-6.5mm}
\DP\ind{(1)}{pert}(-Q_{0}^{2},-Q^{2}) = \ln\biggl(\frac{Q^2}{Q_0^2}\biggr)
+ \frac{4}{\beta_{0}}
\ln\Biggl[\frac{a\ind{(1)}{pert}(Q_{0}^{2})}{a\ind{(1)}{pert}(Q^{2})}\Biggr].
\end{equation}
In these equations $a(Q^2)=\alpha(Q^2)\beta_{0}/(4\pi)$,
\begin{equation}
\alpha\ind{(1)}{pert}(Q^2) =
\frac{4\pi}{\beta_{0}}\,\frac{1}{\ln z}, \qquad z=\frac{Q^2}{\Lambda^2}
\end{equation}
is the one--loop perturbative running coupling, and
\begin{equation}
\label{NAIC}
\alpha\ind{(1)}{an}(Q^2) =
\frac{4\pi}{\beta_{0}}\,\frac{z-1}{z\,\ln z}
\end{equation}
stands for the one--loop infrared enhanced analytic running
coupling~\cite{PRD6264, Review}, which was independently rediscovered in
Refs.~\cite{Schrempp, RCProsperi}.

The perturbative approximation of~$\Pi(q^2)$~(\ref{PPert1L}) contains
infrared unphysical singularities, that makes it inapplicable at low
energies. The expressions~(\ref{P_DPT2}) and~(\ref{PAPT1L}) contain no
unphysical singularities, but their infrared behavior is quite different.
Specifically, the APT prediction~(\ref{PAPT1L}) diverges at~$Q^2 \to 0$
(that makes it also inapplicable at low energies), whereas the~DPT
expression~(\ref{P_DPT2}) vanishes in the infrared limit and proves to be
applicable in the entire energy range.

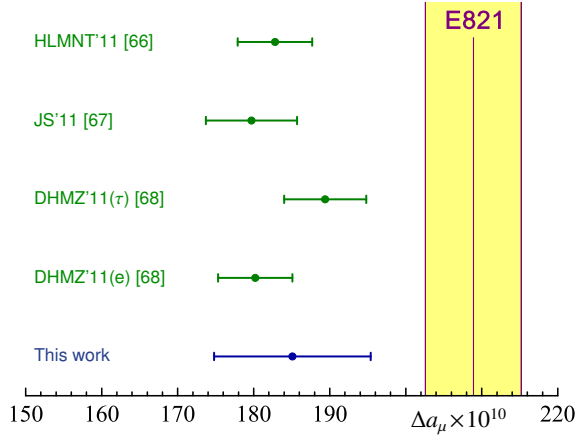
\begin{figure}[t]
\centerline{\includegraphics[width=75mm,clip]{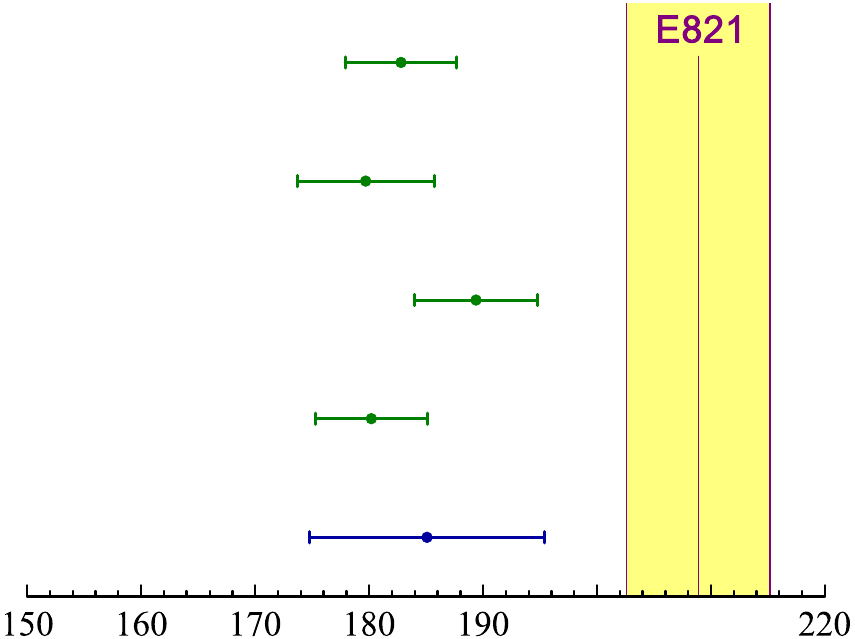}%
\hbox to 0pt {\hss%
\unitlength=1mm
\begin{picture}(74,53.5)
\put(52.075,-0.5){\small$\Delta a_{\mu}\!\times\! 10^{10}$}
\put(2.5,49.9){\txt{HLMNT'11~\cite{HLMNT11}}{G4}}
\put(2.5,39.5){\txt{JS'11~\cite{JS11}}{G4}}
\put(2.5,29.1){\txt{DHMZ'11($\tau$)~\cite{DHMZ11}}{G4}}
\put(2.5,18.7){\txt{DHMZ'11(e)~\cite{DHMZ11}}{G4}}
\put(2.5,8.3){\txt{This work}{RB4}}
\end{picture}}}
\unitlength=1pt
\caption{The subtracted muon anomalous magnetic moment ($\Delta a_{\mu} =
a_{\mu} - a_{0}$, $\,a_{0} = 11659 \times 10^{-7}$): theoretical
evaluations (circles) and experimental measurement (shaded band).}
\label{Plot:Amu}
\end{figure}

\medskip

The persisting few standard deviations discrepancy between the
experimental measurements~\cite{MuonExp1, MuonExp2} and theoretical
evaluations~\cite{MuonRev1, MuonRev2} of the muon anomalous magnetic
moment~\mbox{$a_{\mu} = (g_{\mu}-2)/2$} makes the latter a challenging
issue of particle physics. The uncertainty of theoretical estimation
of~$a_{\mu}$ is largely dominated by the leading--order hadronic
contribution~\cite{Raf72}
\begin{equation}
\label{AmuHVP}
a_{\mu}^{\mbox{\tiny HLO}} =
\frac{1}{3} \biggl(\frac{\alpha}{\pi}\biggr)^{\!2}
\!\int_{0}^{1}\!(1-x)
\bar{\Pi}\biggl(\!m_{\mu}^{2}\,\frac{x^2}{1-x}\!\biggr) dx,
\end{equation}
which involves the integration of~$\Pi(q^2)$ over the range inaccessible
within perturbation theory.

The DPT expression for~$\Pi(q^2)$~(\ref{P_DPT}) contains no unphysical
singularities and enables one to perform the integration in
Eq.~(\ref{AmuHVP}) without invoking experimental data on~$R$--ratio, that
eventually results~in~\cite{JPG42}
\begin{equation}
\label{AmuHLO_DPT}
a_{\mu}^{\mbox{\tiny HLO}} = (696.1 \pm 9.5) \times 10^{-10}.
\end{equation}
This equation corresponds to the four--loop level and the quoted error
accounts for the uncertainties of the parameters entering
Eq.~(\ref{AmuHVP}), their values being taken from
Ref.~\cite{PDG2012CODATA2012}. The obtained estimation~(\ref{AmuHLO_DPT})
appears to be in a good agreement with its recent
assessments~\cite{HLMNT11, JS11, DHMZ11}.

The complete muon anomalous magnetic moment comprises the QED
contribution~\cite{AmuQED}, the electroweak contribution~\cite{AmuEW}, as
well as the higher--order~\cite{HLMNT11} and
light--by--light~\cite{AmuHlbl} hadronic contributions, that, together
with $a_{\mu}^{\mbox{\tiny HLO}}$~(\ref{AmuHLO_DPT}) lead to $a_{\mu} =
(11659185.1 \pm 10.3) \times 10^{-10}$, see Ref.~\cite{JPG42}. The
discrepancy between this value and experimental measurement
$a_{\mu}^{\mbox{\scriptsize exp}} = (11659208.9 \pm 6.3) \times
10^{-10}$~\cite{MuonExp2, MuonExp3} corresponds to two~standard
deviations. As one can infer from Fig.~\ref{Plot:Amu}, the obtained
$a_{\mu}$ conforms with its recent evaluations~\cite{HLMNT11, JS11,
DHMZ11}.

\begin{figure}[t]
\centerline{\includegraphics[width=75mm,clip]{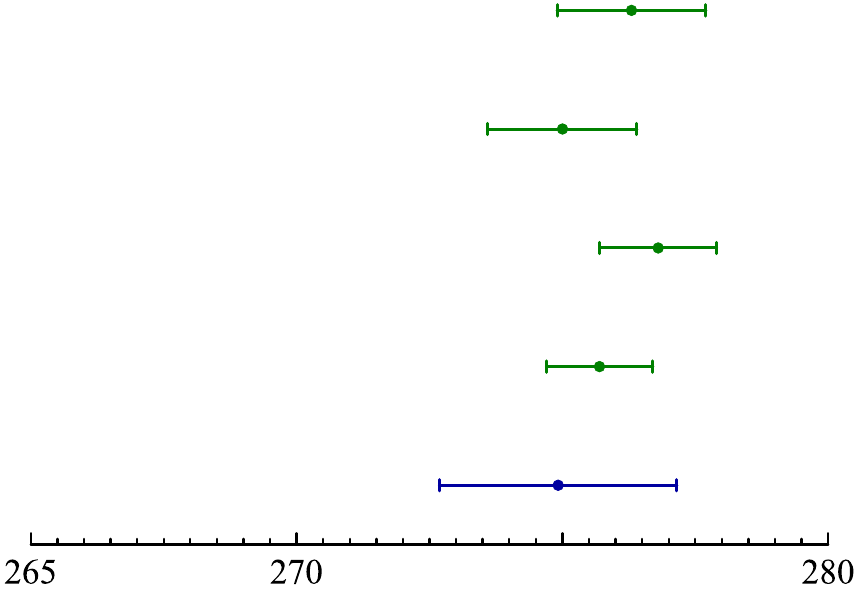}%
\hbox to 0pt {\hss%
\unitlength=1mm
\begin{picture}(74,53.5)
\put(40.5,-0.5){\small$\Delta\alpha\ind{(5)}{had}(M\inds{2}{Z})\!\times\! 10^{4}$}
\put(3,50.1){\txt{HLMNT'11~\cite{HLMNT11}}{G4}}
\put(3,39.7){\txt{J'11~\cite{J11}}{G4}}
\put(3,29.3){\txt{DHMZ'12($\tau$)~\cite{DHMZ11}}{G4}}
\put(3,18.9){\txt{DHMZ'12(e)~\cite{DHMZ11}}{G4}}
\put(3,8.5){\txt{This work}{RB4}}
\end{picture}}}
\unitlength=1pt
\caption{Theoretical evaluations of hadronic contribution to the shift
of electromagnetic fine structure constant at the scale of $Z$~boson
mass.}
\label{Plot:AHMZ}
\end{figure}
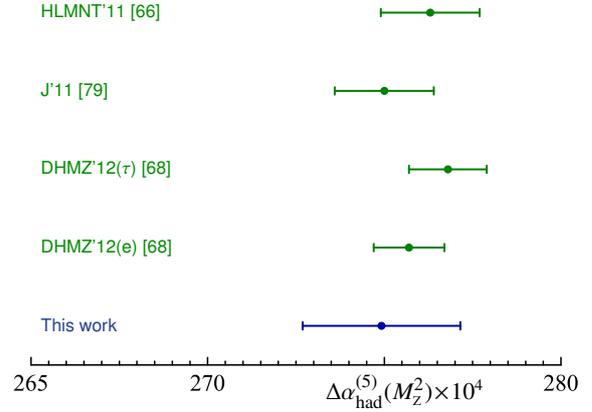

Another observable of our interest is the electromagnetic running coupling
\begin{equation}
\label{RC_QED}
\alpha\ind{}{em}(q^2) = \alpha \bigl[1 - \AL{}(q^2) - \AH{}(q^2)\bigr]^{-1},
\end{equation}
which plays a key role in a variety of issues of precision particle
physics. The leptonic contribution~$\AL{}(q^2)$ to Eq.~(\ref{RC_QED}) can
be calculated by making use of perturbation theory~\cite{AEMLep}, whereas
the hadronic contribution
\begin{equation}
\label{AMH}
\AMH{}(q^2) = - \frac{\alpha}{3\pi}\,q^2
\,\,\mathcal{P}\!\!\int_{m^2}^{\infty}\!\!
\frac{R(s)}{s-q^2}\,\frac{d\,s}{s}
\end{equation}
involves the integration over the low--energy range and constitutes the
prevalent source of the uncertainty of~$\alpha\ind{}{em}(q^2)$, see, e.g.,
Refs.~\cite{HLMNT11, Passera}.

To evaluate the five--flavor hadronic contribution to the shift of the
electromagnetic fine structure constant at the scale of $Z$~boson mass
within~DPT we shall follow the very same lines as above, that eventually
yields~\cite{JPG42}
\begin{equation}
\label{AMH_DPT}
\AMH{(5)}(M\inds{2}{Z}) = (274.9 \pm 2.2) \times 10^{-4}.
\end{equation}
This equation corresponds to the four--loop level and the quoted error
accounts for the uncertainties of the parameters entering Eq.~(\ref{AMH}),
their values being taken from Ref.~\cite{PDG2012CODATA2012}. The obtained
estimation of~$\AMH{(5)}(M\inds{2}{Z})$~(\ref{AMH_DPT}) is in a good
agreement with its recent evaluations~\cite{HLMNT11, DHMZ11, J11},
see~Fig.~\ref{Plot:AHMZ}. At~the same time, Eq.~(\ref{AMH_DPT}) together
with leptonic~\cite{AEMLep} and top~quark~\cite{AEMtop} contributions
results in $\alpha\ind{-1}{em}(M\inds{2}{Z}) = 128.962 \pm 0.030$, that
also conforms with recent assessments of this quantity~\cite{HLMNT11,
DHMZ11, J11}, see Ref.~\cite{JPG42} for the details.

\smallskip

The author is grateful to G.~Bali, R.~Kaminski, K.~Miura, M.~Passera,
J.~Portoles, P.~Roig, and H.~Wittig for the stimulating discussions and
useful comments.

\end{document}